# Algorithms and Approaches of Proxy Signature: A Survey

MANIK LAL DAS
Dhirubhai Ambani Institute of Information and Communication Technology
Gandhinagar, India.
Email: maniklal@gmail.com

ASHUTOSH SAXENA
Infosys Technologies Limited
Hyderabad, India.
Email: Ashutosh_Saxena01@infosys.com

DEEPAK B. PHATAK
Indian Institute of Technology, Bombay
Mumbai, India.
Email: dbp@it.iitb.ac.in

September 11, 2018

**Abstract**

Numerous research studies have been investigated on proxy signatures over the last decade. This survey reviews the research progress on proxy signatures, analyzes a few notable proposals, and provides an overall remark of these proposals.

## 1 Introduction

Digital signature is a cryptographic means through which the authenticity, data integrity and senders non-repudiation can be verified. Typically, digital signature of a document is a piece of information encrypted by the signer's secret key. Numerous researches have shown significant contributions to this field using various cryptographic primitives [55]. However, there are many practical environments where digital signatures do not possess specific requirements, and thereby digital signatures appear in several other forms *viz.* proxy signatures, multi signatures, blind signatures, etc. For example, a manager of a company wants to go for a long trip. He would need a proxy agent, to whom he would delegate his signing capability, and thereafter the proxy agent would sign the documents on behalf of the manager.

The concept of proxy signature was recorded in 1989 [24]; however, the cryptographic treatment on proxy signature was geared up after the scheme by Mambo et al [53] in 1996. They first classified the proxy signature on the basis of delegation, namely *full delegation*, *partial delegation* and *delegation by warrant*, and presented a well devised scheme. In *full delegation*, an original signer gives his secret key to a proxy signer and the proxy signer signs document using original signer's secret key. The drawback of proxy signature with *full delegation* is that the absence of a distinguishability between original signer and proxy signer.

In *partial delegation*, the original signer derives a proxy key from his secret key and hands it over to the proxy signer as a delegation capability. In this case, the proxy signer can misuse the delegation capability, because *partial delegation* can not restrict the proxy signer's signing capability. The weaknesses of *full* and *partial delegations* are eliminated by *partial delegation with warrant*. A warrant explicitly states the signers' identity, delegation period and the qualification of the message on which the proxy signer can sign, etc. Another important requirement of a proxy signature is that the revocation of delegation capability (i.e., proxy revocation). The proxy revocation is essential for the situation where original signer key is compromised or any misuse of the delegation capability is noticed. It may so happen that the original signer wants to terminate his delegation capability before its expiry. Though, Mambo et al's [53] scheme presented an informative idea on proxy signatures and its various features; however, the scheme allows unlimited delegation, i.e., the proxy signer can sign any message because the original signer delegation provides unlimited signing capability to the proxy signer. The unlimited signing capability allows proxy signer to misuse the delegation capability.

In 1997, Kim et al [40] proposed a scheme by restricting proxy signer signing capability using the concept of *partial delegation with warrant*. Subsequently, Zhang [87], [88] proposed threshold and non-repudiable proxy signature schemes. Ghodosi and Pieprzyk [26] analyzed the shortcomings of [87] and the same is also noticed by Lee et al [44]. Petersen and Horster [62] proposed another notion, called self-certified keys under different trust levels and used them for creation of delegation capability, delegated signatures and proxy signatures. However, the work in [47] and [43] showed that Pertersen-Horster's scheme is insecure.

In 1999, Okamoto et al [59], for the first time, proposed proxy signature based on RSA signature scheme, but they considered the proxy unprotected notion. In the same year, Sun proposed two schemes [72], [76] on threshold proxy signatures. Then, Lee and Kim [46] proposed a strong proxy signature scheme. Later, Viswanathan et al [80] proposed a scheme for controlled environments.

In 2000, Sun [73], [74] proposed a multi-proxy signature scheme and a time-stamped proxy signature, respectively. Hwang et al [37] presented a non-repudiable threshold proxy signature scheme with known signers. Subsequently, Yi et al [86] proposed a proxy multi-signature scheme.

In 2001, Romao and da Silva [65] proposed secure mobile agent with proxy certificates. Lee et al [47], [48] proposed two proxy signature schemes and highlighted a few applications. However, Wang et al. [82] noticed that Lee et al's scheme [48] is not secure. Park and Lee [60] proposed another scheme for mobile communications.

In 2002, Shum and Wei [70] proposed a proxy signature scheme with proxy signer privacy protection, but the scheme's insecurity was noticed in [75].

Year 2003 saw a very impressive list of publications demonstrating a vigourous interest in the proxy signature study. In this year, a number of new schemes and improvements have been proposed [79], [36], [49], [7], [31], [32], [43], [82], [15], [13], [68], [51], [38] [39], [42], [34], [17], [18], [50], [90], [89], [1]. However, most of the schemes observed the insecurity of previously proposed schemes and proposed an improved one, which was subsequently broken by others.

In 2004, a few interesting schemes [33], [19], [81], [14], [69], [83], [52], [35], [77], [78], [85], [91], have been proposed.

In 2005, Lee and Lee [45] further addressed the security weaknesses of [70]. Later, Das et al. [20] proposed a pairing-based proxy signature scheme with revocation.

In this survey, we review several notable proxy signature schemes categorizing them into different constructions based on their security assumptions. The organization of the survey is as follows. In the next section, we give a mathematical background for general readership. Section 3 discusses the security properties of proxy signature. Section 4 details the constructions of various proxy signatures. Section 5 reviews some of the notable schemes. Finally, Section 6 gives an overall remarks on our observations.

# 2 Preliminaries

## 2.1 Discrete Logarithm Problem

The discrete logarithm is the inverse of discrete exponentiation in a finite cyclic group. Suppose $G$ be a finite multiplicative cyclic group of order a large prime $q$. Let $g$ be a generator of $G$. Then every element $y$ of $G$ can be written in the form $y = g^k$ for some integer $k$. The discrete logarithm of $y$ is $k$ and is written as $\log_g y = k \bmod q$.
In 1976, Diffie and Hellman [21] revolutionized the cryptography and proposed a key exchange protocol without using secure channel, where the security of the protocol relies on discrete logarithm problem (DLP). Since then, several key exchange [11], public key encryption [12], [55], and signature schemes [22], [27], [71] have been proposed in which the security assumptions trust on the hardness of the DLP.

**The Schnorr signature scheme**: The scheme $\mathcal{S}_{sch}$ is based on DLP and works as follows:
`Setup` ($\mathcal{SP}_{dlp}$): Inputs $1^k$; and outputs **params-dlp**. The **params-dlp** consists of primes $q$ and $l$ such that $2^{k-1} \leq q < 2^k$, an element $g \in \mathbb{Z}_q^*$ of order $l$ that divides $q-1$, and a hash function $h : \{0,1\}^* \rightarrow \mathbb{Z}_l$.
`KeyGen` ($\mathcal{KG}_{dlp}$): The users agree on a group $G$ (multiplicative group of integers modulo $q$ for some prime $q$ with generator $g$ of prime order $l$ in which the DLP is hard. The user chooses a secret key $x \in \mathbb{Z}_l$. The public key is generated as $y = g^x \bmod q$.
In other words, user public key $\leftarrow \mathcal{KG}_{dlp}$(**params-dlp**, user secret key),
i.e. $y \leftarrow \mathcal{KG}_{dlp}$(**params-dlp**, $x$)
`Sign` ($\mathcal{S}_{dlp}$): To sign a message $m$, choose a random $t \in \mathbb{Z}_l$ and compute $r = g^t \bmod q$. Compute $c = h(m, r)$ and $\sigma = (t - xc) \bmod l$. The signature of $m$ is $(\sigma, c)$.
In other words, $\sigma \leftarrow \mathcal{S}_{dlp}$(**params-dlp**, $(t, r)$, $x$, $m$).
`Verify` ($\mathcal{V}_{dlp}$): Compute $r' = g^\sigma y^c \pmod{q}$ and $c' = h(m, r')$. If $c' = c$ then the signature is valid. In other words, **Result**$\leftarrow \mathcal{V}_{dlp}$(**params-dlp**, $y$, $\sigma$, $m$), where **Result** $\in \{Valid, Invalid\}$.
The Schnorr's signature scheme is proven secure [71] under the assumption that DLP is hard.

## 2.2 Bilinear Pairings

Bilinear pairings were first introduced to elliptic curve cryptography for destructive methods like the MOV reduction [54]. With the help of Weil pairing, the authors of [54] showed a way to reduce the DLP on supersingular elliptic curves to the DLP of an extension of the underlying finite field. Later, Frey and Ruck [23] extended the attack to general elliptic curves with the Tate pairing. However, the Weil pairing and the Tate pairing can also be used as a constructive tool for cryptography [9], [10], [16], [30].

Suppose $G_1$ is an additive cyclic group generated by $P$ of prime order $q$, and $G_2$ is a multiplicative cyclic group of the same order. A map $\hat{e} : G_1 \times G_1 \rightarrow G_2$ is called a bilinear mapping if it satisfies the following properties:
- Bilinear: $\hat{e}(aP, bQ) = \hat{e}(P, Q)^{ab}$ for all $P, Q \in G_1$ and $a, b \in \mathbb{Z}_q^*$;
- Non-degenerate: There exist $P, Q \in G_1$ such that $\hat{e}(P, Q) \neq 1$;
- Computable: There exist efficient algorithm to compute $\hat{e}(P, Q)$ for all $P, Q \in G_1$.

In general, $G_1$ is a group of points on an elliptic curve and $G_2$ is a multiplicative subgroup of a finite field.

**Computational Problems**

Definition 1. Discrete Logarithm Problem (DLP) : Given $Q, R \in G_1$, find an integer $x \in \mathbb{Z}_q^*$ such that $R = xQ$.

Definition 2. Decisional Diffie-Hellman Problem (DDHP) : Given $(P, aP, bP, cP)$ for $a, b, c \in \mathbb{Z}_q^*$, determine whether $c \equiv ab \bmod q$.

The advantage of any PPT algorithm $\mathcal{A}$ in solving DDHP in $G_1$ is defined as $\texttt{Adv}_{\mathcal{A},G_1}^{DDH}$ $= [\texttt{Prob}[\mathcal{A}(P, aP, bP, cP) = 1]–\texttt{Prob}[\mathcal{A}(P, aP, bP, abP) = 1]\colon\ a, b \in \mathbb{Z}_q^*]$. For every PPT algorithm $\mathcal{A}$, $\texttt{Adv}_{\mathcal{A},G_1}^{DDH}$ is negligible.

Definition 3. Computational Diffie-Hellman Problem (CDHP) : Given $(P, aP, bP)$ for $a, b \in \mathbb{Z}_q^*$, compute $abP$.

The advantage of any probabilistic polynomial-time (PPT) algorithm $\mathcal{A}$ in solving CDHP in $G_1$, is defined as $\texttt{Adv}_{\mathcal{A},G_1}^{CDH} = \texttt{Prob}[\mathcal{A}(P, aP, bP, abP) = 1 : a, b \in \mathbb{Z}_q^*]$. For every PPT algorithm $\mathcal{A}$, $\texttt{Adv}_{\mathcal{A},G_1}^{CDH}$ is negligible.

Definition 4. Gap Diffie-Hellman (GDH) group: A prime order group $G_1$ is a GDH group if there exists an efficient polynomial-time algorithm which solves the DDHP in $G_1$ and there is no PPT algorithm which solves the CDHP with non-negligible probability of success. The domains of bilinear pairings provide examples of GDH groups. The MOV reduction [54] provides a method to solve DDHP in $G_1$, whereas there is no known efficient algorithm for CDHP in $G_1$.

Definition 5. Bilinear Diffie-Hellman Problem (BDHP) : Given $(P, aP, bP, cP)$ for $a, b, c \in \mathbb{Z}_q^*$, compute $\hat{e}(P, P)^{abc}$.

Definition 6. Weak Diffie-Hellman Problem (WDHP) : Given $(P, Q, aP)$ for $a \in \mathbb{Z}_q^*$, compute $aQ$.

Since most of our discussions on pairing-based proxy signatures are surrounded by Hess's signature scheme [30], we discuss the scheme as follows.

**The Hess signature scheme**: The scheme $\mathcal{S}_{hess}$ is based on CDHP and works as follows:
`Setup`$(\mathcal{SP}_{cdhp})$: It takes $1^k$ and master-key $s$ as input; and outputs **params-cdhp**. The **params-cdhp** includes groups $G_1$, $G_2$ of order prime $q$; a generator $P \in G_1$; a bilinear map $\hat{e} : G_1 \times G_1 \to G_2$; map-to-point $H : \{0, 1\}^* \to G_1$, hash function $h : \{0, 1\}^* \times \{0, 1\}^* \to \mathbb{Z}_q^*$ and public key of a trusted party, say Key Generation Center (KGC) $(Pub_{KGC} = sP)$. The KGC keeps $s$ secret.

`KeyGen` $(\mathcal{KG}_{cdhp})$: It takes **params-cdhp**, user (with identity ID) public key $Pub_{ID} = H(ID)$ as input; outputs user secret key $S_{ID} = sPub_{ID}$, i.e., $S_{ID} \leftarrow \mathcal{KG}_{cdhp}($**params-cdhp**, $Pub_{ID})$.

`Sign` $(\mathcal{S}_{cdhp})$: To sign a message $m$, the signer chooses an arbitrary $P_1 \in G_1$, picks a random $t \in \mathbb{Z}_q^*$ and computes

$r = \hat{e}(P_1, P)^t$.

$c = h(m, r)$.

$\sigma = cS_{ID} + tP_1$.

The signature of $m$ is the tuple $(c, \sigma)$.
In other words, $\sigma \leftarrow \mathcal{S}_{cdhp}(\textbf{params-cdhp}, (t, r, c), S_{ID}, m)$
`Verify` ($\mathcal{V}_{cdhp}$): The signature $(c, \sigma)$ is verified by the following checking:
Compute $r' = \hat{e}(\sigma, P) \cdot \hat{e}(H(ID), -Pub_{KGC})^c$. Accept the signature if $c = h(m, r')$.
In other words, **Result**$\leftarrow \mathcal{V}_{cdhp}(\textbf{params-cdhp}, Pub_{ID}, Pub_{KGC}, \sigma, (c, m))$, where **Result** $\in \{Valid, Invalid\}$.
The Hess's signature scheme is proven secure against existential forgery on adaptive chosen-message attack under the assumption that CDHP is hard.

### 2.3 Integer Factorization Problem

The integer factorization (also known as prime decomposition) problem (IFP) is: *Input* a large positive integer; *output* it as a product of prime numbers. The problem holds a strong security assumption in cryptography, complexity theory, and quantum computers. Based on IFP, the most widely use scheme for public key encryption and signature is RSA [64]. There are many algorithms, protocols, and products where security relies on IFP.

**The RSA signature scheme**: The signature scheme $\mathcal{S}_{rsa}$ works as follows:
`Setup`($\mathcal{SP}_{rsa}$): Inputs $1^k$, secret large primes $p$, $q$; and outputs **params-rsa**. The **params-rsa** consists of a public modulus $N = pq$ and hash function $h : \{0,1\}^* \rightarrow \mathbb{Z}_N$.
`KeyGen` ($\mathcal{KG}_{rsa}$): Choose public key $e$ such that $1 < e < \phi(N)$ which is co-prime to $\phi(N)$, where $\phi(N) = (p-1)(q-1)$. Compute secret key $d$ such that $de \equiv 1 \bmod \phi(N)$. Procedurally, $d \leftarrow \mathcal{KG}_{rsa}(\textbf{params-rsa}, e)$.
`Sign` ($\mathcal{S}_{rsa}$) : To sign a message $m$, compute $\sigma = h(m)^d \bmod N$. The signature of a message $m$ is the tuple $(m, \sigma)$.
In other words, $\sigma \leftarrow \mathcal{S}_{rsa}(\textbf{params-rsa}, d, m)$.
`Verify` ($\mathcal{V}_{rsa}$) : Compute $m' = \sigma^e \bmod N$. The signature is valid if $m' = h(m)$.
In other words, **Result**$\leftarrow \mathcal{V}_{rsa}(\textbf{params-rsa}, e, \sigma, m)$, **Result** $\in \{Valid, Invalid\}$.

## 3 Security Properties of a Proxy Signature

Desirable security properties of proxy signatures have evolved from the introduction of proxy signature. A widely accepted list of required properties is given below:

- Strong unforgeability: A designated proxy signer can create a valid proxy signature on behalf of the original signer. But the original signer and other third parties cannot create a valid proxy signature.

- Strong identifiability: Anyone can determine the identity of corresponding proxy signer from the proxy signature.

- Strong undeniability: Once a proxy signer creates a valid proxy signature on behalf of the original signer, he cannot deny the signature creation.

- Verifiability: The verifier can be convinced of the signers' agreement from the proxy signature.

- Distinguishability: Proxy signatures are distinguishable from the normal signatures by everyone.

- Secrecy: The original signer secret key cannot be derived from any information, such as the shares of the proxy key, proxy signatures, etc.

- Prevention of misuse: The proxy signer cannot use the proxy key for other purposes than it is made for. That is, he cannot sign message with the proxy key that have not been defined in the warrant. If he does so, he will be identified explicitly from the warrant.

## 3.1 Classification of Proxy Signature

According to the nature of delegation capability, proxy signature can be classified as proxy-unprotected, proxy-protected and threshold notions. This differentiation is important in practical applications, since it enables proxy signature schemes to avoid potential disputes between the original signer and proxy signer.

### 3.1.1 Proxy-unprotected notion

The scenario exists when an original signer gives his signing rights (full delegation with warrant) to a proxy signer. The original signer sends a signed warrant to the proxy signer, who then uses this information to generate proxy signatures by executing a standard signature scheme. When a proxy signature is sent, the recipient checks its validity according to the corresponding standard signature verification process. As the proxy signer does not append his secret key on top of the received delegation, a dishonest original signer can sign the message and later claim that the signature was created by the proxy signer. This type of proxy signature primarily lacks strong unforgeability property.

### 3.1.2 Proxy-protected notion

This is the scenario when the proxy signer uses his secret key to safeguard him from the original signer's forgery. In this case, the original signer sends a signed warrant to the proxy signer, who then uses it to construct a proxy key by appending his secret key. With the proxy key, the proxy signer can generate proxy signatures by executing a standard signature scheme. When a proxy signature is sent, the recipient first computes the proxy public key from some public information, and then checks its validity according to the corresponding standard signature verification process. By this technique, neither the original signer frames proxy signer nor the proxy signer frames original signer.

### 3.1.3 Threshold notion

In a threshold proxy signature, the proxy key is shared by a group of $n$ proxy signers. In order to produce a valid proxy signature on a given message $m$, individual proxy signer produce his partial signature on that message, and then combines them into a full proxy signature on message $m$.
In a $(t, n)$ threshold proxy signature scheme, the original signer delegates his signing capability to a proxy group of $n$ members. Any $t$ or more proxy signers of the group can cooperatively issue a proxy signature on behalf of the original signer, but $(t-1)$ or less proxy signers cannot forge a signature.

# 4 Models of Proxy Signature

Based on the security assumptions on various proxy signature schemes, we categorize the existing schemes into four different constructions: DLP-based proxy signature, RSA-based proxy signature, ECDSA-based proxy signature, and Pairing-based proxy signature. The schemes consist of delegation capability generation, delegation capability verification, proxy key generation, proxy signature generation and proxy signature verification.

## 4.1 DLP-based Proxy Signature

The participants involved in the model are:
- an original signer, who delegates his signing capability to a proxy signer.
- a proxy signer, who signs the message on behalf of the original signer.
- a verifier, who verifies the proxy signature and decides to accept or reject.
- a trusted party who certifies the public key.

**DLP-based Proxy Signature Model**: An original signer selects a secret key $x_o$ and computes his public key $y_o$ as

$y_o \leftarrow \mathcal{KG}_{dlp}(\textbf{params-dlp}, x_o)$.

A proxy signer selects a secret key $x_p$ and computes his public key $y_p$ as

$y_p \leftarrow \mathcal{KG}_{dlp}(\textbf{params-dlp}, x_p)$.

Delegation capability generation: It takes **params-dlp**, original signer chosen parameters $(k_o, r_o)$, original signer secret key $x_o$, a warrant $\omega$ as input; and outputs signature $\sigma_o$ on $\omega$,

Procedurally, $\sigma_o \leftarrow \mathcal{S}_{dlp}(\textbf{params-dlp}, (k_o, r_o), x_o, \omega)$

Delegation capability verification: It takes **params-dlp**, $y_o$, $\omega$, $\sigma_o$ as input; and outputs **Result**, where **Result**$\in \{Valid, Invalid\}$, i.e., **Result**$\leftarrow \mathcal{V}_{dlp}(\textbf{params-dlp}, y_o, \sigma_o, \omega)$.

Proxy key generation($\texttt{PKeyGen}_{dlp}$): It takes **params-dlp**, $\sigma_o$, $x_p$ and random number as input; and outputs proxy key $\rho_p$. Typically, the proxy signer uses simple arithmetic operation to form a proxy key $\rho_p = y_o\sigma_o + x_p y_p \bmod q$.

Procedurally, $\rho_p \leftarrow \texttt{PKeyGen}_{dlp}(\textbf{params-dlp}, \sigma_o, x_p, \textit{pub-params}^1)$

Proxy signature generation: It takes **params-dlp**, proxy key $\rho_p$ and message $m$ as input; outputs signature $\sigma_p$ on $m$, i.e, $\sigma_p \leftarrow \mathcal{S}_{dlp}(\textbf{params-dlp}, \rho_p, m)$

Proxy signature verification: It takes **params-dlp**, $y_o$, $y_p$, $m$ and $\sigma_p$ as input; outputs **Result**, i.e., **Result**$\leftarrow \mathcal{V}_{dlp}(\textbf{params-dlp}, (y_o, y_p), \sigma_p, m)$.

## 4.2 RSA-based proxy signature

The participants involved in the model are:
- an original signer, who delegates his signing capability to a proxy signer.
- a proxy signer, who signs the message on behalf of the original signer.
- a verifier, who verifies the proxy signature and decides to accept or reject.
- a trusted party who certifies the public key.

**RSA-based Proxy Signature Model**: An original signer selects a public key $y_o$ and computes his secret key $x_o$ as

[1] *pub-params* include signers' public keys, random numbers, warrant, etc.

$x_o \leftarrow \mathcal{KG}_{rsa}(\textbf{params-rsa},\ y_o)$.
A proxy signer selects a public key $y_p$ and computes his secret key $x_p$ as
$x_p \leftarrow \mathcal{KG}_{rsa}(\textbf{params-rsa},\ y_p)$.

Delegation capability generation: It takes **params-rsa**, $x_o$, a warrant $\omega$ as input; and outputs signature $\sigma_o$ on $\omega$, Procedurally, $\sigma_o \leftarrow \mathcal{S}_{rsa}(\textbf{params-rsa},\ x_o,\ \omega)$

Delegation capability verification: It takes **params-rsa**, $y_o$, $\omega$, $\sigma_o$ as input; and outputs **Result**. That is, **Result**$\leftarrow \mathcal{V}_{rsa}(\textbf{params-rsa},\ y_o,\ \sigma_o,\ \omega)$, where **Result** $\in \{Valid, Invalid\}$.

Proxy signature generation: It takes **params-rsa**, $\sigma_o$, $x_p$ and message $m$ as input; outputs signature $\sigma_p$ on $m$, i.e, $\sigma_p \leftarrow \mathcal{S}_{rsa}(\textbf{params-rsa},\ x_p,\ (\sigma_o, m))$

Proxy signature verification: It takes **params-rsa**, $y_o$, $y_p$, $m$ and $\sigma_p$ as input; and outputs **Result**, i.e., **Result**$\leftarrow \mathcal{V}_{rsa}(\textbf{params-rsa},\ (y_o, y_p),\ \sigma_p, m)$.

### 4.3 Pairing-based proxy signature

The participants involved in the model are:
- an original signer, who delegates his signing capability to a proxy signer.
- a proxy signer, who signs the message on behalf of the original signer.
- a verifier, who verifies the proxy signature and decides to accept or reject.
- a trusted party who issues user secret key.

**Pairing-based Proxy Signature Model**: The original signer generates his public key $y_o = H(ID_o)$, and computes secret key $x_o \leftarrow \mathcal{KG}_{cdhp}(\textbf{params-cdhp},\ y_o)$, where $ID_o$ is original signer's identity.
The proxy signer generates his public key $y_p = H(ID_p)$, and computes secret key
$x_p \leftarrow \mathcal{KG}_{cdhp}(\textbf{params-cdhp},\ y_p)$, where $ID_p$ is proxy signer's identity.

Delegation capability generation: It takes **params-cdhp**, $x_o$ and a warrant $\omega$ as input; outputs signature $\sigma_o$ on $\omega$, i.e., $\sigma_o \leftarrow \mathcal{S}_{cdhp}(\textbf{params-cdhp},\ (k_o, r_o, c_o),\ x_o,\ \omega)$.

Delegation capability verification: It takes **params-cdhp**, $y_o$, $\omega$ and $\sigma_o$ as input; and outputs **Result**, i.e., **Result**$\leftarrow \mathcal{V}_{cdhp}(\textbf{params-cdhp},\ (y_o,\ Pub_{KGC}),\ \sigma_o,\ (c_o, \omega))$, where **Result** $\in \{Valid, Invalid\}$.

Proxy key generation: It takes **params-cdhp**, $\sigma_o$, $x_p$ and random number as input; and outputs proxy key $\rho_p \leftarrow \texttt{PKeyGen}_{cdhp}(\textbf{params-cdhp},\ \sigma_o,\ (\textit{user-params}^2),\ x_p)$.

Proxy signature generation: It takes **params-cdhp**, proxy key $\rho_p$ and message $m$ as input; outputs signature $\sigma_p$ on $m$, i.e., $\sigma_p \leftarrow \mathcal{S}_{cdhp}(\textbf{params-cdhp},\ (k_p, r_p),\ \rho_p,\ m)$.

Proxy signature verification: It takes **params-cdhp**, $y_o$, $y_p$, $m$ and $\sigma_p$ as input; outputs **Result**, i.e., **Result**$\leftarrow \mathcal{V}_{cdhp}(\textbf{params-cdhp},\ (y_o, y_p, Pub_{KGC}),\ \sigma_p,\ (c_p, m, \omega))$.

[2] The *user-params* includes signers' public keys, random numbers, warrant, etc.

# 5 Review of Some Notable Proxy Signatures

## 5.1 DLP-based proxy signature schemes

| *Conventions and notation for DLP-based proxy signature schemes* | |
|---|---|
| Alice | Original signer |
| Bob | Proxy signer |
| $q$ | A large prime |
| $\mathbb{Z}_q$ | Set of integers modulo $q$ |
| $\mathbb{Z}_q^*$ | Multiplicative group of $\mathbb{Z}_q$ |
| $g$ | Generator of large order in $\mathbb{Z}_q^*$ |
| $x_o, x_p$ | Secret key of Alice and Bob, respectively |
| $y_o$ | Public key of Alice, $y_o = g^{x_o} \bmod q$ |
| $y_p$ | Public key of Bob, $y_p = g^{x_p} \bmod q$ |
| $\omega$ | A warrant |
| $h(.)$ | A collision-resistant one-way hash function |

### 5.1.1 Mambo, Usuda and Okamoto [53]

First classified the proxy signature on the basis of the degree of delegation, and proposed a well devised scheme. In the scheme both proxy unprotected and proxy protected notions are envisaged. As we are more focused on proxy-protected scheme, here we discuss the proxy-protected notion.

Assumption: DLP is hard.

Alice selects a secret key $x_o$ and generates public key $y_o \leftarrow \mathcal{KG}_{dlp}(\textbf{params-dlp}, x_o)$.
Bob selects a secret key $x_p$ and generates public key $y_p \leftarrow \mathcal{KG}_{dlp}(\textbf{params-dlp}, x_p)$.
Delegation capability generation: Alice chooses a random number $k_o \in \mathbb{Z}_{q-1}^*$ and computes $r_o = g^{k_o} \bmod q$. Alice computes $\sigma_o \leftarrow \mathcal{S}_{dlp}(\textbf{params-dlp}, (k_o, r_o), x_o)$.
Delegation capability verification: Bob accepts $\sigma_o$ if and only if

$$Valid \leftarrow \mathcal{V}_{dlp}(\textbf{params-dlp}, y_o, \sigma_o).$$

Proxy key generation: Bob computes proxy key $\rho_p \leftarrow \texttt{PKeyGen}_{dlp}(\textbf{params-dlp}, \sigma_o, x_p, \textit{pub-params})$.
Proxy signature generation: The proxy signature on message $m$ is computed as

$$\sigma_p \leftarrow \mathcal{S}_{dlp}(\textbf{params-dlp}, (k_p, r_p), \rho_p, m).$$

Proxy signature verification: The verifier accepts the proxy signature if and only if

$$Valid \leftarrow \mathcal{V}_{dlp}(\textbf{params-dlp}, (y_o, y_p), \sigma_p, m).$$

Security: The underlying security of the scheme is based on the hardness of DLP. However, the scheme has two weaknesses. Firstly, unlimited delegation, i.e., Bob can sign any message on behalf of Alice because Alice has delegated unlimited signing rights to Bob. Secondly, proxy transfer, i.e., If Bob transfers Alice's delegation power to any other party who can sign any message on behalf of Alice. In other words, the scheme does not satisfy the prevention of misuse security property.

#### 5.1.2 Kim, Park and Won [40]

Proposed proxy signature for partial delegation with warrant and proxy signature for threshold delegation.

Assumption: DLP is hard.

Scheme for partial delegation with warrant (PDW):
Alice chooses a secret key $x_o$ and generates public key $y_o \leftarrow \mathcal{KG}_{dlp}(\textbf{params-dlp}, x_o)$.
Bob chooses a secret key $x_p$ and generates public key $y_p \leftarrow \mathcal{KG}_{dlp}(\textbf{params-dlp}, x_p)$.
Delegation capability generation: Alice chooses a random number $k_o \in \mathbb{Z}^*_{q-1}$ and computes $r_o = g^{k_o} \bmod q$. Then, Alice computes $\sigma_o \leftarrow \mathcal{S}_{dlp}(\textbf{params-dlp}, (k_o, r_o), x_o, \omega)$.
Delegation capability verification: Bob accepts $\sigma_o$ if and only if

$$Valid \leftarrow \mathcal{V}_{dlp}(\textbf{params-dlp}, y_o, \sigma_o, \omega).$$

Proxy key generation: Bob computes proxy key $\rho_p \leftarrow \texttt{PKeyGen}_{dlp}(\textbf{params-dlp}, \sigma_o, x_p, \textit{pub-params})$.
Proxy signature generation: The proxy signature on message $m$ is computed as

$$\sigma_p \leftarrow \mathcal{S}_{dlp}(\textbf{params-dlp}, (k_p, r_p), \rho_p, m).$$

Proxy signature verification: The verifier accepts the proxy signature if and only if

$$Valid \leftarrow \mathcal{V}_{dlp}(\textbf{params-dlp}, (y_o, y_p), \sigma_p, (\omega, m)).$$

Scheme for Threshold Delegation: In the threshold delegation, Alice sends her delegation to a proxy group so that to sign a message the proxy signer's power is shared.
Alice chooses a secret key $x_o$ and generates public key $y_o \leftarrow \mathcal{KG}_{dlp}(\textbf{params-dlp}, x_o)$.
Each proxy signer acts as a dealer with a random secret $u$, chooses a random polynomial such that $f(x) = u + a_1 x + \cdots + a_{t-1} x^{t-1} \bmod q-1$. The proxy group keys are generated as follows:
Public keys: $g^u \bmod q$, $g^{a_1} \bmod q$, $\cdots$, $g^{a_{t-1}} \bmod q$.
Secret keys: $x_{p,i} = u + a_1 i + \cdots + a_{t-1} i^{t-1}$; $i = 1, 2, \cdots, t$.
Delegation capability generation: Alice chooses a random number $k_o \in \mathbb{Z}^*_{q-1}$ and computes $r_o = g^{k_o} \bmod q$. Then Alice computes $\sigma_o \leftarrow \mathcal{S}_{dlp}(\textbf{params-dlp}, (k_o, r_o), x_o, \omega)$.
Proxy Sharing: To share $\sigma_o$ in a threshold manner with threshold $t$, Alice chooses random $b_j \in \mathbb{Z}_{q-1}; j = 1, 2, \cdots, t-1$, and publishes the values $B_j = g^{b_j}, j = 1, 2, \cdots, t-1$. Then, she computes the proxy share $\sigma_i$ as $\sigma_i = f'(i) = \sigma_o + b_1 i + \cdots + b_{t-1} i^{t-1}$.
Delegation capability verification: Each proxy signer accepts $\sigma_i$ if and only if

$$g^{\sigma_i} = y_o^{h(\omega, r_o)} r_o \cdot \prod_{j=1}^{t-1} B_j^{(i^j)} \bmod q.$$

Proxy key generation: Each proxy signer computes proxy key as

$$\rho_{p,i} \leftarrow \texttt{PKeyGen}_{dlp}(\textbf{params-dlp}, \sigma_i, x_{p,i}, \textit{pub-params}).$$

Proxy signature generation: To sign a message $m$, each proxy signer computes $v = h(l, m)$, where $l = g^r \bmod q$ ($r$ is secret to the proxy signer). Then, the proxy signer computes $\lambda_i = s_i + \sigma_i v \bmod q-1$ and reveals $\lambda_i$, where $s_i = f(i) = r + a_1 i + \cdots + a_{t-1} i^{i-1}$. On validating $\lambda_i$, each proxy signer computes $\sigma$ satisfying $\sigma = r + \sigma_o v = f(0) + f'(0) v \bmod q-1$ by applying Lagrange formula to $\lambda_i$. The proxy signature on $m$ is the tuple $(\omega, r_o, m, \sigma, v)$.
Proxy signature verification: The verifier checks whether $v' = g^\sigma \cdot (y_o \cdot y_p)^{h(\omega, r_o)} r_o)^{-v} \bmod q$, and then whether $v = h(v', m)$.

Security: To the best of our knowledge the proposed PDW scheme is still unbroken. However, the intuition in this scheme that using warrant does not require proxy revocation is not correct. There are many situations where proxy revocation is a must though warrant explicitly states the validity and restricts the message signing. Sun et al [76] showed that the above threshold delegation approach is not secure.

#### 5.1.3 Zhang [87]

Proposed threshold and non-repudiable proxy signatures, where both the original signer and proxy signer can not falsely deny their signature.

Assumption: DLP is hard.

Alice chooses a secret key $x_o$ and generates public key $y_o \leftarrow \mathcal{KG}_{dlp}(\textbf{params-dlp}, x_o)$.
Assume that there is a group of $n$ proxy signers $p_i$, $i = 1, \cdots, n$.
Proxy key generation: Alice picks $k_o \in \mathbb{Z}_{q-1}$, computes $R = g^{k_o} \bmod q$, and broadcasts $R$. The proxy signer randomly selects $\alpha_i \in \mathbb{Z}_{q-1}$, computes $y_{p_i} = g^{\alpha_i} R \bmod q$, checks whether $y_{p_i} \in \mathbb{Z}^*_{q-1}$ and if it holds then broadcasts $y_{p_i}$.
Alice computes $\hat{R} = \prod_{i=1}^{n} y_{p_i}$, and $\hat{s} = n^{-1}\hat{R}x_o + k \bmod q-1$ and broadcasts $\hat{s}$. Then, each proxy signer computes $\hat{R} = \prod_{i=1}^{n} y_{p_i}, \sigma_{p_i} = \hat{s} + \alpha_i \bmod q-1$, and checks if the equality holds: $g^{\hat{s}} = y^{n^{-1}\hat{R}} R \bmod q$. If it holds, the proxy signer accepts $\sigma_{p_i}$ as a valid proxy share.
The threshold proxy signature and verification are done in similar approaches as in the schemes [29], [25].

Security: Lee et al.'s [44] and [76] pointed out some weaknesses in Zhang's threshold proxy signatures [87]. Later, some additional attacks are commented in [26].

#### 5.1.4 Petersen and Horster [62]

Proposed a scheme for self-certified keys issuance under different trust level and used them for delegation of signing rights and delegated signatures, proxy signatures.

Assumption: DLP is hard.

Alice picks a secret key $x_o$ and generates public key $y_o \leftarrow \mathcal{KG}_{dlp}(\textbf{params-dlp}, x_o)$.
Bob picks a secret key $x_p$ and generates public key $y_p \leftarrow \mathcal{KG}_{dlp}(\textbf{params-dlp}, x_p)$.
Delegation capability generation: Alice picks a random number $k_o \in \mathbb{Z}^*_{q-1}$, computes $r_o = g^{k_o} \bmod q$. Then, Alice computes $\sigma_o \leftarrow \mathcal{S}_{dlp}(\textbf{params-dlp}, (k_o, r_o), x_o, ProxyID)$.
Delegation capability verification: Bob accepts $\sigma_o$ if and only if

$$Valid \leftarrow \mathcal{V}_{dlp}(\textbf{params-dlp}, y_o, \sigma_o, ProxyID).$$

Proxy key generation: Bob computes proxy key $\rho_p \leftarrow \texttt{PKeyGen}_{dlp}(\textbf{params-dlp}, \sigma_o, x_p, \textit{public-parameters})$.
Proxy signature generation: The proxy signature on message $m$ is computed as

$$\sigma_p \leftarrow \mathcal{S}_{dlp}(\textbf{params-dlp}, \rho_p, (m, ProxyID)).$$

Proxy signature verification : The verifier accepts the proxy signature if and only if

$$Valid \leftarrow \mathcal{V}_{dlp}(\textbf{params-dlp}, (y_o, y_p), \sigma_p, (m, ProxyID)).$$

Security: The scheme has three weaknesses. Firstly, the proxy signer can deny his signature creation later because a proxy signature does not contain any authentic information of proxy signer. Secondly, the proxy signer gets a proxy key pair $(x_p, y_p)$ from the original signer. He can deny his signature by showing that the proxy signature is created by the original signer with the name of him. Thirdly, the original signer sends the signing rights to a proxy signer without any agreement or warrant. The original signer can argue that the proxy signature is not valid for the concerned message. Moreover, the schemes in [47], [43] further pointed out some more weaknesses of this scheme.

#### 5.1.5 Sun, Lee and Hwang [76]

Proposed a $(t, n)$ threshold proxy signature based on Zhang's threshold scheme [87].

Assumption: DLP is hard.

Alice picks a secret key $x_o$ and generates public key $y_o \leftarrow \mathcal{KG}_{dlp}(\textbf{params-dlp}, x_o)$.
Assume that there is a group of $n$ proxy signers $p_i$, $i = 1, \cdots, n$. Each proxy signer $p_i$ has a secret key $x_{p_i} \in \mathbb{Z}_{q-1}$ and a public key $y_{p_i} \leftarrow \mathcal{KG}_{dlp}(\textbf{params-dlp}, x_{p_i})$.
Proxy generation: Alice picks $k_o \in \mathbb{Z}_{q-1}$, computes $R = g^{k_o} \bmod q$, and broadcasts $R$. Then each proxy signer randomly selects $\alpha_i \in \mathbb{Z}_{q-1}$, computes $r_{p_i} = g^{\alpha_i} R \bmod q$.
Alice computes $\hat{R} = \prod_{i=1}^{n} r_{p_i} \bmod q$, and $\hat{s} = n^{-1} x_o h(\hat{R}, PGID) + k_o \bmod q-1$ and broadcasts $\hat{s}$, where $PGID$ is the proxy group identity that records the proxy status, the event mark of the proxy share generation, the expiration time of the delegation, the identities of original signer and proxy signers. After validating $\hat{s}$, each proxy signer performs a $(t, n)$ verifiable threshold secret sharing scheme [61], and acts as a dealer to distribute proxy sub-shares to other $n-1$ proxy signers for generating their valid proxy shares. Each proxy signer $p_i$ selects a $(t-1)$-degree polynomial $f_i(x) = s_i + a_{i,1}x + a_{i,2}x^2 + \cdots + a_{i,t-1}x^{t-1} \bmod q-1$, where $s_i = \hat{s} + \alpha_i + x_{p_i} h(\hat{R}, PGID) \bmod q-1$. Then $p_i$ sends the proxy sub-share $f_i(j)$ to proxy signer $p_j$(for $1 \leq j \leq n$ and $j \neq i$) vis a secure channel. In addition, $p_i$ also broadcasts $g^{a_{i,1}}, \cdots, g^{a_{i,t-1}}$.
After validating all $f_j(i)$, $p_i$ computes $x'_i = \sum_{j=1}^{n} f_j(i) \bmod q-1$ as his proxy share. Let $f(x) = \sum_{j=1}^{n} f_j(x) \bmod q-1$. This proxy share can be written as $x'_i = f(i)$ and will be used for generating proxy signatures. The shared secret key is regarded as $f(0) = \hat{n}s + \sum_{i=1}^{n} \alpha_i + \sum_{i=1}^{n} x_{p_i} h(\hat{R}, PGID) = \sum_{i=1}^{n} (\alpha_i + k_o) + \sum_{i=0}^{n} x_{p_i} h(\hat{R}, PGID) \bmod q-1$.
Proxy signature generation: Each participant proxy signer $p_i$ $(1 \leq i \leq t)$ performs a $(t, t)$ verifiable secret sharing scheme by randomly choosing a $(t-1)$-degree polynomial $f'_i(x) = \sum_{j=0}^{t-1} a'_{i,j} x^j \bmod q-1$ and broadcasts $c'_{i,j} = g^{a'_{i,j}} \bmod q$ for $j = 0, 1 \cdots, t-1$. Then $p_i$ computes $f'_i(j)$ and sends it to $p_j$ via a secure channel for $1 \leq j \leq n$ and $j \neq i$. After validating each $f'_i(j)$, each participant proxy signer $p_i$ computes $x''_i = f(i) = \sum_{j=1}^{t} f'_j(i) \bmod q-1$, where $f'(x) = \sum_{j=1}^{t} f'_j(x) \bmod q-1$ and $Y = \prod_{k=1}^{t} c'_{k,0} \bmod q$. Finally, each $p_i$ computes and broadcasts $T_i = x'_i h(m) + x''_i Y \bmod q-1$.
On validating $T_i$, each $p_i$ computes $T = f(0)h(m) + f'(0)Y \bmod q-1$ from $T_j$ by applying Lagrange's interpolating polynomial, where $m$ is the message. The proxy signature on message $m$ is the tuple $(\hat{R}, PGID, Y, T)$.
Proxy signature verification: The verifier accepts the proxy signature if and only if

$$g^T = \left( \left( y_o \prod_{i=1}^{n} y_i \right)^{h(\hat{R}, PGID)} \hat{R} \right)^{h(m)} Y^Y \bmod q.$$

Security: Hsu et al. [34] and Shao [69] noticed that Sun et al.'s scheme is not secure, it lacks coalition attack.

#### 5.1.6 Lee, Kim and Kim [48]

Proposed a scheme in which a mobile agent is constructed using non-designated proxy signature which represents both the original signer's (customer) and the proxy signer's (remote server) signatures. The work provides Schnorr-based and RSA-based constructions for secure mobile agent. Here, we give the Schnorr-based construction, the RSA-based construction is given in the next section.

Assumption: DLP is hard.

Alice chooses a secret key $x_o$ and generates public key $y_o \leftarrow \mathcal{KG}_{dlp}(\textbf{params-dlp}, x_o)$.
Bob chooses a secret key $x_p$ and generates public key $y_p \leftarrow \mathcal{KG}_{dlp}(\textbf{params-dlp}, x_p)$.
Delegation capability generation: Alice chooses a random number $k_o \in \mathbb{Z}^*_{q-1}$ and computes $r_o = g^{k_o} \bmod q$. Then she computes $\sigma_o \leftarrow \mathcal{S}_{dlp}(\textbf{params-dlp}, (k_o, r_o), x_o, \omega)$.
Delegation capability verification: Bob accepts $\sigma_o$ if and only if

$$Valid \leftarrow \mathcal{V}_{dlp}(\textbf{params-dlp}, y_o, \sigma_o, \omega).$$

Proxy key generation: Bob computes proxy key $\rho_p \leftarrow \texttt{PKeyGen}_{dlp}(\textbf{params-dlp}, \sigma_o, x_p, \textit{public-parameters})$.
Proxy signature generation: The proxy signature on message $m$ is computed as

$$\sigma_p \leftarrow \mathcal{S}_{dlp}(\textbf{params-dlp}, (k_p, r_p), \rho_p, m).$$

Proxy signature verification: The verifier accepts the proxy signature if and only if

$$Valid \leftarrow \mathcal{V}_{dlp}(\textbf{params-dlp}, (y_o, y_p), \sigma_p, (m, \omega)).$$

Security: Wang et al. [82] showed that the scheme is insecure against transferring, forgery attacks.

#### 5.1.7 Boldyreva, Palacio and Warinschi [7]

First proposed a formal security notion for proxy signature. At the same time, they proposed a provable secure scheme, called *triple* Schnorr proxy signature scheme, which is an enhanced version of the scheme [40].

Assumption: DLP is hard.

Alice chooses a secret key $x_o$ and generates public key $y_o \leftarrow \mathcal{KG}_{dlp}(\textbf{params-dlp}, x_o)$.
Bob chooses a secret key $x_p$ and generates public key $y_p \leftarrow \mathcal{KG}_{dlp}(\textbf{params-dlp}, x_p)$.
Delegation capability generation: Alice chooses a random number $k_o \in \mathbb{Z}^*_{q-1}$ and computes $r_o = g^{k_o} \bmod q$. Then computes $\sigma_o \leftarrow \mathcal{S}_{dlp}(\textbf{params-dlp}, (k_o, r_o), x_o, (\omega, y_o, y_p))$.
Delegation capability verification: Bob accepts $\sigma_o$ if and only if

$$Valid \leftarrow \mathcal{V}_{dlp}(\textbf{params-dlp}, y_o, \sigma_o, (\omega, y_o, y_p)).$$

Proxy key generation: Bob computes proxy key $\rho_p \leftarrow \texttt{PKeyGen}_{dlp}(\textbf{params-dlp}, \sigma_o, x_p, \textit{pub-params})$.
Proxy signature generation: The proxy signature on message $m$ is computed as

$$\sigma_p \leftarrow \mathcal{S}_{dlp}(\textbf{params-dlp}, (k_p, r_p, y_o, y_p), \rho_p, m)).$$

Proxy signature verification: The verifier accepts the proxy signature if and only if

$$Valid \leftarrow \mathcal{V}_{dlp}(\textbf{params-dlp}, (y_o, y_p), \sigma_p, m)).$$

Security: The scheme is based on Kim et al's PDW scheme [40]. However, they did not consider warrant in the proxy signing phase which leads unlimited delegation, that is, the scheme suffers from delegation misuse.

#### 5.1.8 Li, Tzeng and Hwang [49]

Proposed a generalized version $(t_1/n_1 - t_2/n_2)$ proxy signature scheme . The $(t_1/n_1 - t_2/n_2)$ proxy signature scheme allows $t_1$ out of $n_1$ original signers to delegate their signing capability to a designated proxy group of $t_2$ out of $n_2$ proxy signers. The proxy group of proxy signers can cooperatively generate the proxy signature on behalf of the original group. Any verifier can verify the proxy signature on the message with the knowledge of the identities of the actual original signers and the actual proxy signers.

Assumption: DLP is hard.

Let the scheme consists of $n_1$ original signers and $n_2$ proxy signers.
For $i = 1, 2, \cdots, n_1$, the original signer chooses a secret key $x_{o_i}$ and generates public key $y_{o_i}$ $\leftarrow \mathcal{KG}_{dlp}(\textbf{params-dlp}, x_{o_i})$.
For $j = 1, 2, \cdots, n_2$, the proxy signer chooses a secret key $x_{p_j}$ and generates public key $y_{p_j}$ $\leftarrow \mathcal{KG}_{dlp}(\textbf{params-dlp}, x_{p_j})$.
Delegation capability generation: For $i = 1, 2, \cdots, t_1$ $(< n_1)$, the original signer chooses a random number $k_{o_i} \in \mathbb{Z}^*_{q-1}$ and computes $r_{o_i} = g^{k_{o_i}} \bmod q$. Then the original signer computes $\sigma_{o_i} \leftarrow \mathcal{S}_{dlp}(\textbf{params-dlp}, (k_{o_i}, r_{o_i}), x_{o_i}, \omega)$. A designated clerk (any one of the original signers) verifies the individual proxy shares as whether

$$Valid \leftarrow \mathcal{V}_{dlp}(\textbf{params-dlp}, y_{o_i}, \sigma_{o_i}, \omega).$$

If it holds, the clerk combines the individual proxy shares as $\sigma_o = \sum_{i=1}^{t_1} \sigma_{o_i} \bmod q-1$. The final proxy share is the tuple $(\omega, K, \sigma_o)$, where $K = \prod_{i=1}^{t_1} r_{o_i}$.

Delegation capability verification: Bob accepts $\sigma_o$ if and only if
$Valid \leftarrow \mathcal{V}_{dlp}(\textbf{params-dlp}, (K, y_o), \sigma_o, \omega)$.
Proxy signature generation: Each proxy signer selects a random integer $k_{p_j} \in \mathbb{Z}^*_{q-1}$, computes $r_{p_j} = g^{k_{p_j}} \bmod q$, and broadcast $r_{p_j}$. Then, each proxy signer computes $R = \prod_{j=1}^{t_2} r_{p_j} \bmod q$ and $\sigma_{p_j} = k_{p_j} R + (\sigma_o t_2^{-1} + x_{o_i} y_{o_i})^{h(m,R,ProxyID)} \bmod q-1$, where $t_2$ is the threshold value of the proxy signers group.
A designated clerk verifies the individual proxy signature as whether
$g^{\sigma_{p_j}} = r_{p_j}^R((K^K \prod_{i=1}^{t_1} y_{o_i}^{y_{o_i} h(\omega,K)})^{t_2^{-1}} y_{p_j}^{y_{p_j}})^{h(m,R,ProxyID)} \bmod q$. If it does, the clerk combines the individual proxy signature of $m$ as $\sigma = \sum_{j=1}^{t_2} \sigma_{p_j} \bmod q-1$. The proxy signature of $m$ is $(\omega, K, m, R, \sigma)$.
Proxy signature verification: A verifier checks the validity of the proxy signature on $m$ whether $g^\sigma = R^R(K^K \prod_{i=1}^{t_1} y_{o_i}^{y_{o_i} h(\omega,K)} \prod_{j=1}^{t_2} y_{p_j}^{y_{p_j}})^{h(m,R,ProxyID)} \bmod q$.
Security: The security analysis of the scheme is not rigorous.

#### 5.1.9 Herranz and Saez [33]

Proposed a distributed proxy signature scheme. The scheme extended the work in [7] to the scenario of fully distributed proxy signature schemes.

Assumption: DLP is hard.

Generation of keys: Let $E = \{P^{(1)}, P^{((2)}, \cdots, P^{(n)}\}$ be a distributed entity formed by $n$ participants. There is an *access structure* $\Gamma \subset 2^E$, which is formed by those subsets of participants which are authorized to perform the secret task. The access structure must be monotone increasing; that is, if $A_1 \in \Gamma$ is authorized and $A_1 \subset A_2 \subset E$, then $A_2$ must be authorized. The joint generation of discrete logarithm keys is as follows:
Each participant $P^{(l)} \in E$ obtains a secret value $x^{(l)} \in \mathbb{Z}_{q-1}$. The values $\{x^{(l)}\}_{P^{(l)} \in E}$ form a sharing of the secret key $x \in \mathbb{Z}_{q-1}$, according to some linear secret sharing scheme realizing the access structure $\Gamma$. The corresponding public key $y = g^x \bmod q$ is made public, along with other values (commitments) which ensure the robustness of the protocol.
The fully distributed *triple* Schnorr proxy signature scheme is generated in a similar way of the Boldyreva et al's scheme [7].

Security: The scheme is as secure as Kim et al's PDW scheme [40].

#### 5.1.10 Malkin, Obana and Yung [52]

Presented a formal model for fully hierarchical proxy signatures with warrant that supports chains of several levels of delegation.

Assumption: DLP is hard.

The signers' selects secret key $x_i$ and computes public key $y_i \leftarrow \mathcal{KG}_{dlp}(\textbf{params-dlp}, x_i)$.
Delegation capability generation: This is an interactive process between the designator and proxy signer. It takes public keys of a designator $y_{i_{L-1}}$ and a proxy signer $y_{i_L}$, the signing key of which the designator delegates its signing right (i.e., the signing key is either a signing key $x_{i_{L-1}}$ or a proxy key $\sigma_{i_o \cdots \to i_{L-1}}$ depending on whether $i_{L-1}$ is original signer or proxy signer), a warrant up to previous delegation $W_{L-1}$ and a warrant $\omega_L$ set in current delegation as inputs; outputs delegation rights.
Proxy key generation: It takes public keys of a designator $y_{i_{L-1}}$ and a proxy signer $y_{i_l}$, the secret key of the proxy signer $x_{i_L}$ as inputs and outputs a proxy key $\sigma_{i_o \cdots \to i_L}$ and a warrant $\omega$.
Proxy signature generation : The proxy signature on message $m$ is computed as

$$\sigma_p \leftarrow \mathcal{S}_{dlp}(\textbf{params-dlp}, \sigma_{i_o \cdots \to i_L}, (m, \omega)).$$

Proxy signature verification : The verifier accepts the proxy signature if and only if

$$Valid \leftarrow \mathcal{V}_{dlp}(\textbf{params-dlp}, y_{i_o}, \sigma_p, (m, \omega)).$$

Security: The scheme formalizes a model of fully hierarchical proxy signature, which is a probably secure model to the best of our knowledge.

### 5.2 RSA-based Proxy Signature

#### 5.2.1 Okamoto, Tada and Okamoto [59]

Proposed a scheme that reduces the computation and storage cost during the protocol execution, and the protocol is suitable for implementation in smart card.

Assumption: IFP is hard and smart card is tamper resistant.

Alice chooses a public key $y_o$ and generates secret key $x_o \leftarrow \mathcal{KG}_{rsa}(\textbf{params-rsa}, y_o)$.
Delegation capability generation: Alice computes $\sigma_o \leftarrow \mathcal{S}_{rsa}(\textbf{params-rsa}, x_o, (\omega, I_p))$ where $I_p$ denote the limit of money which she can spend.
Delegation capability verification: Bob accepts $\sigma_o$ if and only if

$$Valid \leftarrow \mathcal{V}_{rsa}(\textbf{params-rsa}, y_o, \sigma_o, (\omega, I_p)).$$

| *Conventions and notation for RSA-based proxy signature schemes* | |
|---|---|
| Alice | Original signer |
| Bob | Proxy signer |
| $N_o, N_p$ | RSA Modulus for Alice and Bob, respectively |
| $y_o$ | Public key of Alice, where $1 < y_o < \phi(N_o)$ |
| $y_p$ | Public key of Bob, where $1 < y_p < \phi(N_p)$ |
| $x_o$ | Secret key of Alice, where $x_o y_o \equiv 1 \bmod \phi(N_o)$ |
| $x_p$ | Secret key of Bob, where $x_p y_p \equiv 1 \bmod \phi(N_p)$ |
| $\omega$ | A warrant |
| $h(.)$ | A collision-resistant one-way hash function |

Proxy signature generation: To sign a message $m$, Bob generates a random number $k_p \in \mathbb{Z}_N^*$, and computes

$r = g^{k_p h(m)} \sigma_o \bmod N_o$

$s = g^{-y_o k_p} \bmod N_o$.

The proxy signature of message $m$ is the tuple $(m, (r, s), I_p)$.
Proxy signature verification: The verifier checks whether $(ID_p, \omega) = h(I_p) r^{y_o} s^{h(m)} \bmod N_o$. If it does, the verifier accepts it as a valid proxy signature. Otherwise, rejects it.

Security: The scheme has a weak security as it is designed as a proxy-unprotected scheme, where Alice can frame Bob by signing the message and later claim that Bob has signed the message.

### 5.2.2 Lee, Kim and Kim [48]

A mobile agent is constructed in the scheme using non-designated proxy signature which represents both the original signer's (customer) and the proxy signer's (remote server) signatures.

Assumption: IFP is hard.

Alice chooses a public key $y_o$ and generates secret key $x_o \leftarrow \mathcal{KG}_{rsa}(\textbf{params-rsa}, y_o)$.
The mobile agent (proxy signer) chooses a public key $y_p$ and generates secret key

$x_p \leftarrow \mathcal{KG}_{rsa}(\textbf{params-rsa}, y_p)$.

Delegation capability generation: Alice creates $\sigma_o \leftarrow \mathcal{S}_{rsa}(\textbf{params-rsa}, x_o, (AliceID, req))$, where $req$ is the customer requirements for purchase such as price range, date, delivery requirements, etc.
Delegation capability verification: The mobile agent accepts $\sigma_o$ if and only if

$Valid \leftarrow \mathcal{V}_{rsa}(\textbf{params-rsa}, y_o, \sigma_o, (AliceID, req))$.

Proxy signature generation: Let $BID$ be the agent's bid information which conforms to $req$. The agent (remote server) tries to sell the product to Alice. The remote server computes $x = h(AliceID, req, AgentID, BID)^{x_p} \bmod N_p$, $y = h(AliceID, req)^x \bmod N_o$ and $z = \sigma_o^x \bmod N_o$. Then sends the tuple $(AliceID, req, AgentID, BID, x, y, z)$ to the mobile agent and the agent will get back to Alice with this tuple as a receipt of the purchase.
Proxy signature verification: Alice receives $(AliceID, req, AgentID, BID, x, y, z)$ from the mobile agent, then she can verify the validity of the purchase by the following:

- Whether $h(AliceID, req, AgentID, BID) = x^{y_p} \bmod N_p$.
- Whether $y = h(AliceID, req)^x \bmod N_o$.
- Whether $y = z^{y_o} \bmod N_o$.

- Whether $BID \in \{req\}$.

Security: Wang et al. [82] showed that the scheme is insecure and inefficient.

#### 5.2.3 Shao [68]

Proposed a proxy signature scheme based on the factoring problem, which combines the RSA signature scheme and the Guillou and Quisquater [28] signature scheme.

Assumption: IFP hard and Guillou-Quisquater signature is secure.

Alice chooses a public key $y_o$ and generates secret key $x_o \leftarrow \mathcal{KG}_{rsa}(\textbf{params-rsa}, y_o)$.
Bob chooses a public key $y_p$ and generates secret key $x_p \leftarrow \mathcal{KG}_{rsa}(\textbf{params-rsa}, y_p)$.
Delegation capability generation: Alice computes proxy key $v = h(\omega, ProxyID)^{-x_o} \bmod N_o$, $u = \lfloor v/N_p \rfloor$ and $z = v^{y_p} \bmod N_p$. The delegation is the tuple $(\omega, z, u)$.
Delegation capability verification: Bob recovers $v = u \times N_p + (z^{x_p} \bmod N_p)$.
Proxy signature generation: Let $m$ be the message to be signed by Bob. Bob does the following:

- Randomly chooses an integer $t \in [1, N_o]$ and computes $r = t^{y_o} \bmod N_o$.
- Compute $k = h(m, r)$ and $x = k^{x_p} \bmod N_p$.
- Compute $y = tv^k \bmod N_o$.

The proxy signature on message $m$ is $(m, \omega, x, y, ProxyID)$.
Proxy signature verification: The verifier checks the following:

- Compute $k' = x^{y_p} \bmod N_p$.
- Compute $r' = y^{y_o} h(\omega, ProxyID)^{k'} \bmod N_o$.
- Check whether $k' = h(m, r')$.

Security: The security of the scheme is based on Guillou-Quisquater signature scheme [28]. However, the author did not give the formal security proof.

#### 5.2.4 Das, Saxena and Gulati [19]

Proposed a proxy signature scheme that provides effective proxy revocation mechanism.

Assumption: IFP is hard.

In addition to Alice and Bob, a trusted server (TS) is another participant in the scheme for time stamp issuance.
Alice chooses a public key $y_o$ and generates secret key $x_o \leftarrow \mathcal{KG}_{rsa}(\textbf{params-rsa}, y_o)$.
Bob chooses a public key $y_p$ and generates secret key $x_p \leftarrow \mathcal{KG}_{rsa}(\textbf{params-rsa}, y_p)$.
TS chooses a public key $y_s$ and generates secret key $x_s \leftarrow \mathcal{KG}_{rsa}(\textbf{params-rsa}, y_s)$.
Delegation capability generation: Alice computes the delegation capability $\sigma_o$ as

$$\sigma_o \leftarrow \mathcal{S}_{rsa}(\textbf{params-rsa}, x_o, (\omega, y_p, y_s)).$$

Delegation capability verification: Bob accepts $\sigma_o$ if and only if

$$Valid \leftarrow \mathcal{V}_{rsa}(\textbf{params-rsa}, y_o, \sigma_o, (\omega, y_p, y_s)).$$

Proxy signature generation: Let $m$ be the message to be signed. Bob requests a time stamp to the TS and sends $(R, m, y_o, y_p)$, where $R \leftarrow \mathcal{S}_{rsa}(\textbf{params-rsa}, x_p, (m, \omega, y_o, y_p))$. The TS verifies whether $Valid \leftarrow \mathcal{V}_{rsa}(\textbf{params-rsa}, y_p, R, (m, \omega, y_o, y_p))$. If it holds, the TS ascertain the following conditions are true before the time stamp is issued:

- Alice's public key $y_o$ is not in the public revocation list maintained by the TS.
- $\omega$ is not expired.

Now, TS computes $T_m \leftarrow \mathcal{S}_{rsa}(\textbf{params-rsa}, x_s, (m, \omega, y_o, y_p, T))$, where $T$ denotes a time stamp. Then, TS sends $(T_m, T)$ to Bob over a public channel.
Bob accepts $T$ if and only if $Valid \leftarrow \mathcal{V}_{rsa}(\textbf{params-rsa}, y_s, T_m, (m, \omega, y_o, y_p, T))$.
If it holds, Bob generates proxy signature as $\sigma_p = (h(m, \omega, y_o, y_s, T) \oplus \sigma_o)^{x_p} \bmod N_p$, otherwise rejects the time stamp and makes another request to the TS. The proxy signature of message $m$ is $(m, \omega, T, T_m, \sigma_p)$.
Proxy signature verification: The verifier checks the following to validate a proxy signature:

- Whether $Valid \leftarrow \mathcal{V}_{rsa}(\textbf{params-rsa}, y_s, T_m, (m, \omega, y_o, y_p, T))$. If it holds, the verifier is assured that the time stamp in the signed message is correct. Otherwise, he rejects the signed message.
- Whether $h(\omega, y_p, y_s) = (\sigma_p^{y_p} \bmod N_p \oplus h(m, \omega, y_o, y_s, T))^{y_o} \bmod N_o$. If it holds, he accepts the signed message. Otherwise, he rejects it.

Security: The scheme provides an effective proxy revocation mechanism. The scheme is secure on the assumption that IFP is hard. However, the scheme does not work when $N_p > N_o$, but it is a valid assumption because typically the proxy signer key strength should not be greater than the original signer key strength.

## 5.3 ECDSA-based Proxy Signature

### 5.3.1 Chen, Chung and Huang [13]

Proposed a scheme for proxy multi-signature based on elliptic curve cryptosystem that reduces high computational overheads of Sun's scheme [74].

Assumption: ECDLP is hard.

Let $B = (x_B, y_B)$ be a point in $E(F_q)$ for a large prime $q$, the order of $B$ is assumed as $t$. For each $1 \leq i \leq n$, the original signer secretly selects a random number $1 \leq d_i \leq t-1$ as her private key and computes the corresponding public key $Q_i = d_i \times B = (x_{Q_i}, y_{Q_i})$.
The proxy signer selects a private key $1 \leq d_p \leq t-1$ and computes corresponding public key $Q_p = d_p \times = (x_{Q_p}, y_{Q_p})$.

Delegation capability generation: For each $1 \leq i \leq n$, the original signer $A_i$ selects a random number $1 \leq k_i \leq t-1$, computes $r_i = k_i \times B = (x_{r_i}, y_{r_i})$ and $s_i = x_i \cdot x_{Q_i} \cdot h(\omega, r_i) k_i \bmod t$.

Delegation capability verification and proxy key generation : For each $1 \leq i \leq n$, the proxy signer computes $U_i = (x_{Q_i} \cdot h(\omega, r_i) \bmod t) \times Q_i - s_i \times B = (x_{U_i}, y_{U_i})$ using $(\omega, r_i, s_i)$. If $x_{U_i} = x_{r_i} \bmod t$, the proxy signer accepts $s_i$ as a valid delegation of signing right; otherwise, he rejects it. If the proxy signer validates all $(\omega, r_i, s_i)$ in which $1 \leq i \leq n$, s/he then computes $d = d_p \cdot x_{Q_p} + \sum_{i=1}^{n} s_i \bmod t$ as a valid proxy key.

Proxy signature generation: When the proxy signer signs a message $m$ for $A_1, \cdots, A_n$, he executes the signing operation of a designated signature scheme using the signing key $d$. The resulting proxy signature is the tuple $(m, \sigma_{s_p}(m), r_1, r_2, \cdots, r_n, \omega)$.

Proxy signature verification: The verifier computes the proxy public key $Q$ corresponding to the proxy key $s_p$ for verifying the proxy signature by the designated signature scheme: $Q = x_{Q_p} \times Q_p + (x_{Q_1} \cdot h(\omega, r_1) \bmod t \times Q_1 + \cdots + (x_{Q_n} \cdot h(\omega, r_n) \bmod t \times Q_n(r_1 + \cdots + r_n)$.
With the newly generated proxy public key $Q$, the verifier confirms the validity of $\sigma_{s_p}(m)$ by validating the verification equation of the designated signature scheme.

Security: The authors did not consider any security model for their scheme, instead, a heuristic security analysis is given to safeguard the scheme.

### 5.4 Pairing-based Proxy Signature

| *Conventions and notation for pairings-based proxy signature schemes* | |
|---|---|
| Alice | Original signer |
| Bob | Proxy signer |
| $G_1$ | A cyclic additive group, whose order is prime $q$ |
| $G_2$ | A cyclic multiplicative group of the same order $q$ |
| $P$ | A generator of $G_1$ |
| $\hat{e}$ | A bilinear pairing map |
| $H(\cdot)$ | Map-to-Point function |
| $h(\cdot)$ | Collision-resistant one-way hash function |
| $s$ | PKG's master-key |
| $Pub_{KGC}$ | KGC's public key, $Pub_{KGC} = sP$ |
| $y_o, x_o$ | Alice's public key and secret key, respectively, $y_o = H(ID_o)$ |
| $y_p, x_p$ | Bob's public key and secret key, respectively, $y_p = H(ID_p)$ |

#### 5.4.1 Zhang, Safavi-Naini and Lin [90]

Proposed an ID-based proxy signature based on Hess's ID-based signature.

Assumption: WDHP is hard.

Alice computes her public key $y_o = H(ID_o)$, where $ID_o$ is her identity. Then, Alice obtains her secret key $x_o \leftarrow \mathcal{KG}_{cdhp}(\textbf{params-cdhp}, y_o)$ from KGC.
Bob computes his public key $y_p = H(ID_p)$, where $ID_p$ is his identity. Then, Bob obtains his secret key $x_p \leftarrow \mathcal{KG}_{cdhp}(\textbf{params-cdhp}, y_p)$ from KGC.
Delegation capability generation: Alice computes $\sigma_o \leftarrow \mathcal{S}_{cdhp}(\textbf{params-cdhp}, (k_o, r_o, c_o), x_o, \omega)$.
Delegation capability verification: Bob accepts $\sigma_o$ if and only if

$$Valid \leftarrow \mathcal{V}_{cdhp}(\textbf{params-cdhp}, y_o, \sigma_o, (c_o, \omega)).$$

Proxy key generation: Bob computes proxy key $\rho_p \leftarrow \texttt{PKeyGen}_{cdhp}(\textbf{params-cdhp}, \sigma_o, c_o, x_p)$.
Proxy signature generation: Bob picks a random number $k_p \in \mathbb{Z}_{q-1}^*$, computes
$r_p = \hat{e}(P, P)^{k_p}$, $c_p = h(m, r_p)$ and $\sigma_p \leftarrow \mathcal{S}_{cdhp}(\textbf{params-cdhp}, (k_p, c_p), \rho_p, m)$.
Proxy signature verification: The verifier accepts the proxy signature if and only if

$$Valid \leftarrow \mathcal{V}_{cdhp}(\textbf{params-cdhp}, (y_o, y_p), \sigma_p, (c_p, m, \omega)).$$

Security: The security proof is not rigorous. Only heuristic security analysis is given to safeguard the scheme.

#### 5.4.2 Chen, Zhang and Kim [15]

Proposed a multi-proxy signature scheme, where Alice delegates her signing capability to $l$ proxy signers.

Assumption: CDHP is hard.

Alice computes her public key $y_o = H(ID_o)$, where $ID_o$ is her identity. Then, Alice obtains her secret key $x_o \leftarrow \mathcal{KG}_{cdhp}(\textbf{params-cdhp}, y_o)$ from KGC.
Each proxy signer computes his public key $y_{p_i} = H(ID_{p_i})$, where $ID_{p_i}$ is his identity. Then, Each proxy signer obtains his secret key $x_{p_i} \leftarrow \mathcal{KG}_{cdhp}(\textbf{params-cdhp}, y_{p_i})$ from KGC.
Delegation capability generation: Alice picks a random number $k_o \in \mathbb{Z}^*_{q-1}$, computes $r_o = \hat{e}(P,P)^{k_o}$, $c_o = h(\omega, r_o)$ and $\sigma_o \leftarrow \mathcal{S}_{cdhp}(\textbf{params-cdhp}, (k_o, r_o, c_o), x_o, \omega)$.
Delegation capability verification: Each proxy signer accepts $\sigma_o$ if and only if

$$(k_p, c_p), \leftarrow \mathcal{V}_{cdhp}(\textbf{params-cdhp}, y_o, \sigma_o, (c_o, \omega)).$$

Proxy key generation: Each proxy signer computes proxy key as

$$\rho_{p_i} \leftarrow \texttt{PKeyGen}_{cdhp}(\textbf{params-cdhp}, \sigma_o, c_o, x_{p_i}).$$

Proxy signature generation: Each proxy signer performs the following operations to sign a message $m$:

- Pick randomly $k_{p_i} \in \mathbb{Z}^*_{q-1}$, computes $r_{p_i} = \hat{e}(P,P)^{k_{p_i}}$ and broadcasts $r_{p_i}$ to the remaining $l-1$ proxy signers.

- Compute $r_p = \prod_{i=1}^{l} r_{p_i}$ and $c_p = h(m, r_p)$, $\sigma_{p_i} = c_p \rho_{p_i} + k_{p_i} P$. Then, send $\sigma_{p_i}$ to a designated clerk (one of the proxy signers).

- The clerk verifies the individual proxy signature and computes $\sigma_p = \sum_{i=1}^{l} \sigma_{p_i}$. The proxy signature of message $m$ is the tuple $(m, \omega, r_o, c_p, \sigma_p)$.

Proxy signature verification: The verifier accepts the proxy signature of message $m$ if and only if $c_p = h(m, \hat{e}(\sigma_p, P)(\hat{e}(\sum_{i=1}^{l}(y_o + y_{p_i}, Pub_{KGC})^{h(\omega, r_o)} \cdot r_o^l)^{-c_p}$.
Security: The security proof is not rigorous.

### 5.4.3 Xu, Zhang and Feng [85]

Formalized a notion of security for ID-based proxy signature schemes and presented a proxy signature scheme.

Assumption: CDHP is hard.

Alice computes her public key $y_o = H(ID_o)$, where $ID_o$ is her identity. Then, Alice obtains her secret key $x_o \leftarrow \mathcal{KG}_{cdhp}(\textbf{params-cdhp}, y_o)$ from KGC.
Bob computes his public key $y_p = H(ID_p)$, where $ID_p$ is his identity. Then, Bob obtains his secret key $x_p \leftarrow \mathcal{KG}_{cdhp}(\textbf{params-cdhp}, y_p)$. from KGC.
Delegation capability generation: Alice randomly picks $k_o \in \mathbb{Z}^*_q$, computes $r_o = k_o P$, $C_o = H_o(ID_o, \omega, r_o)$ and $\sigma_o = x_o + k_o C_o$, where $H_o : \{0,1\}^* \times \{0,1\}^* \times G_1 \rightarrow G_1$.
Delegation capability verification: Bob accepts $\sigma_o$ if and only if

$$\hat{e}(\sigma_o, P) = \hat{e}(Pub_{KGC}, y_o)\hat{e}(r_o, C_o).$$

Proxy key generation: Bob computes proxy key $\rho_p = h_1(ID_o, ID_p, \omega, r_o)x_p + \sigma_o$, where $h_1 : \{0,1\}^* \times \{0,1\}^* \times \{0,1\}^* \times G_1 \rightarrow \mathbb{Z}^*_q$.
Proxy signature generation: To sign a message $m$, Bob does the following:
- Picks $k_p \in \mathbb{Z}^*_{q-1}$, computes $r_p = k_p P$ and then puts $C_p = H_o(ID_p, m, r_p)$.
- Computes $\sigma_p = \rho_p + k_p C_p$.

The proxy signature of message $m$ is the tuple $((r_o, r_p), \sigma_p, (\omega, m))$.

Proxy signature verification: The verifier accepts the proxy signature of message $m$ iff $\hat{e}(\sigma_p, P) = \hat{e}(Pub_{KGC}, y_p)^{h_1((ID_o, ID_p, \omega, r_o))}\hat{e}((Pub_{KGC}, y_o)\hat{e}(r_p, C_p)\hat{e}(r_o, C_o)$.

Security: Security of the scheme is based on the CDHP in the random oracle model. The scheme takes large computation cost.

### 5.4.4 Zhang, Safavi-Naini and Susilo [91]

Proposed a proxy signature scheme based on a short signature scheme.

Assumption: CDHP is hard.

Alice computes her public key $y_o = H(ID_o)$, where $ID_o$ is her identity. Then, Alice obtains her secret key $x_o \leftarrow \mathcal{KG}_{cdhp}(\textbf{params-cdhp}, y_o)$ from KGC.
Bob computes his public key $y_p = H(ID_p)$, where $ID_p$ is his identity. Then, Bob obtains his secret key $x_p \leftarrow \mathcal{KG}_{cdhp}(\textbf{params-cdhp}, y_p)$ from KGC.
Delegation capability generation: Alice computes $\sigma_o = (s_o + h(\omega))^{-1}y_p$.
Delegation capability verification: Bob accepts $\sigma_o$ if and only if

$$\hat{e}(h(\omega)P + y_o, \sigma_o) = \hat{e}(P, y_p).$$

Proxy key generation: Bob computes $\rho_p = x_p\sigma_o$.
Proxy signature generation: To sign a message $m$, Bob does the following.
- Chooses a random number $r \in \mathbb{Z}^*_{q-1}$ and computes $U = r \cdot (h(\omega)P + y_o)$.
- Computes $t = H_2(m, U)$ and $\sigma_p = (t + r)^{-1}\rho_p$ , where $H_2 : \{0,1\}^* \times G_1 \to \mathbb{Z}^*_q$.

The proxy signature of message $m$ is $(U, \sigma_p, \omega)$.
Proxy signature verification : The verifier verifies whether

$$\hat{e}(U + H_2(m, U)(h(\omega)P + y_o), \sigma_p) = \hat{e}(y_p, y_p).$$

Security: Security of the scheme is based on CDHP in the random oracle model.

### 5.4.5 Das, Saxena and Phatak [20]

Proposed a proxy signature scheme based on Hess signature scheme that provides effective proxy revocation mechanism and avoids key escrow problem.

Assumption: CDHP is hard.

Alice computes her public key $y_o = H(ID_o)$, where $ID_o$ is her identity. Then, Alice generates her secret key $x_o \leftarrow \mathcal{KG}_{cdhp}(\textbf{params-cdhp}, y_o)$.
Bob computes his public key $y_p = H(ID_p)$, where $ID_p$ is his identity. Then, Bob generates his secret key $x_p \leftarrow \mathcal{KG}_{cdhp}(\textbf{params-cdhp}, y_p)$.
Delegation capability generation: Alice computes $\sigma_o = (s_o + b_oH'(\omega, y_o, y_p)$, and $\psi_o = b_oP$. Here, $b_o$ is secret to the original signer only and $H' : \{0,1\}^* \times G_1 \times G_1 \to G_1$.
Delegation capability verification: Bob accepts $\sigma_o$ if and only if

$\hat{e}(s_o, P) = \hat{e}(\psi_o, H'(\omega, y_o, y_p)) \cdot \hat{e}(y_o, Reg_o)$, where $Reg_o = sb_oP$, registration token published by the KGC.
Proxy key generation: Bob computes $\rho_p = s_o + s_p + b_pH'(\omega, y_o, y_p)$. Here, $b_p$ is secret to the proxy signer only.
Proxy signature generation: To sign a message $m$, Bob does the following.
- Selects a random $r \in \mathbb{Z}^*_q$ and compute $R = rP$.
- Computes $a = h(m, R, y_p)$ and $\psi_p = b_pP$, where $h : \{0,1\}^* \times G_1 \times G_1 \to \{0,1\}^*$.
- Computes $\sigma_p = (r + a)^{-1}\rho_p$.

The proxy signature of message $m$ is $(\omega, m, R, \sigma_p, \psi_o, \psi_p, y_o, y_p)$.

Proxy signature verification : The proxy signature is valid if and only if
$\hat{e}(R + h(m, R, y_p)P, \sigma_p) = \hat{e}(\psi_o + \psi_p, H'(\omega, y_o, y_p)) \cdot \hat{e}(y_o, Reg_o) \cdot \hat{e}(y_p, Reg_p)$.
Security: The scheme is proven secure and does not require secure channel in the key issuance.

# 6 Concluding Remarks

We reviewed a few seminal work on proxy signatures from the different security assumptions. We now compare the reviewed schemes in a tabular manner, highlighting the important features at a glance. In the following, Table-1 depicts the features of DLP-based schemes, Table-2 depicts the features of RSA-based schemes, and Table-3 depicts the features of Pairing-based schemes.

| *Features* → *Schemes* ↓ | *Secure* | *Secure Channel* | *Proxy Revocation* | *Remarks* |
|---|---|---|---|---|
| Mambo et al[53] | No | Yes | Partial | Unlimited delegation and misuse of delegation |
| Kim et al[40] | Yes | No | No | Secure |
| Zhang[87] | No | Yes | No | Insecure |
| Lee et al[48] | No | No | No | Insecure, suffers from Transferring, Forgery attacks |
| Boldyreva et al[7] | Yes | No | No | Based on [40] and formalizes the security notion, but unlimited delegation |
| Li et al[49] | Yes | No | No | No formal security analysis |
| Herranz et al[33] | Yes | No | No | Distributed proxy signature, ideas based on Boldyreva et al[7] |
| Malkin et al[52] | Yes | No | No | Secure, based on [40], hierarchical delegation |

Table 1: Computation time: DLP-based proxy signatures

| *Features* → *Schemes* ↓ | *Secure* | *Secure Channel* | *Proxy Revocation* | *Remarks* |
|---|---|---|---|---|
| Okamoto et al[59] | Yes | Yes | No | Does not provide strong unforgeability and prevention of misuse |
| Lee at al[48] | No | Yes | No | Insecure |
| Shao[68] | No | Yes | No | No formal security proof |
| Das et al[19] | Yes | No | Yes | $N_p < N_o$ |

Table 2: Computation Time : RSA-based proxy signatures

It is observed that many times, a paper typically breaks a previous scheme and proposes a new one, which someone breaks later and, in turn, proposes a new one, and so on. Most of such work, though quite important and useful, essentially provides an incremental advance to the same basic theme.

| *Features* → *Schemes* ↓ | *Key Escrow* | *Secure Channel* | *Proxy Revocation* | *Remarks* |
|---|---|---|---|---|
| Zhang et al[90] | Yes | Yes | No | No formal security proof, suffer from key escrow, need secure channel |
| Chen et al[15] | Yes | Yes | No | No formal security proof, suffer from key escrow, need secure channel |
| Xu et al[85] | Yes | Yes | No | Takes high computation cost, suffer from key escrow, need secure channel |
| Zhang et al[91] | Yes | Yes | No | suffer from key escrow, need secure channel |
| Das et al[20] | No | No | Yes | Secure, no key escrow, no secure channel |

Table 3: Computation Time : Pairing-based proxy signatures

The authors tried to explore whether there are any real implementation of various proposed proxy signatures. They contacted over email individual author(s) of some of the papers, to learn whether their proposed scheme is used in real life scenario? Unfortunately, the responses from several authors indicated that they were not aware of such applications which use their scheme. Some even suggested that, since the authors work with real life banks problems, whether it will possible to try their scheme in such applications.

In conclusion, we believe that the actual deployment of proxy signatures is yet to start in a big way. However, as and when this happens, the research work being carried out will certainly provide practically usable implementations.